\documentclass[pre,twocolumn,showpacs,floatfix]{revtex4}
\usepackage{graphicx}

\def\be{\begin{equation}}
\def\ee{\end{equation}}

\begin{document}
\title{Skewness of probability density functions of fluid particle
acceleration in developed turbulence}
 \author{A.K. Aringazin}
 \email{aringazin@mail.kz}
  \altaffiliation[Also at ]
  {Department of Mechanics and Mathematics, Kazakhstan Division, Moscow State
University, Moscow 119899, Russia.}
 \affiliation{Department of Theoretical Physics, Institute for
Basic Research, Eurasian National University, Astana 473021
Kazakhstan}

\date{5 January 2004}

\begin{abstract}
Within the framework of one-dimensional Laval-Dubrulle-Nazarenko
type model for the Lagrangian acceleration in developed turbulence
studied in the work 
[A.K. Aringazin and M.I. Mazhitov, cond-mat/0305186] we focus on
the effect of correlation between the multiplicative noise and the
additive one which models the relationship between the stretching
and vorticity, and can be seen as a skewness of the probability
density function of some acceleration component. The skewness of
the acceleration distribution in the laboratory frame of reference
should be zero in the ideal case of statistically homogeneous and
isotropic developed turbulent flows but when considering
acceleration component aligned to fluid particle trajectories it
is of much importance in understanding of the cascade picture in
the three-dimensional turbulence related to Kolmogorov four-fifths
law. We illustrate the effect of nonzero cross correlation
parameter $\lambda$. With $\lambda=-0.005$ the transverse ($x$)
acceleration probability density function turns out to be in good
agreement with the recent experimental data by Mordant, Crawford,
and Bodenschatz. In the Random Intensity of Noise (RIN) approach,
we study the conditional probability density function and
conditional mean acceleration assuming the additive noise
intensity to be dependent on velocity fluctutions.
\end{abstract}


\maketitle

\section{Introduction}\label{Sec:Introduction}

Tsallis nonextensive statistics \cite{Tsallis} inspired
approach~\cite{Johal,Aringazin,Beck3} was recently
used~\cite{Beck,Beck4} to describe Lagrangian acceleration of
fluid particle in developed turbulence; see
also~\cite{Wilk,Beck2,Reynolds}. In Ref.~\cite{Aringazin5} we
reviewed some refinements of this
approach~\cite{Aringazin2,Aringazin3,Aringazin4}.

Review and critical analysis of the applications of various recent
nonextensive statistics based models to the turbulence have been
made by Gotoh and Kraichnan~\cite{Kraichnan0305040}. An emphasis
was made that some models lack justification of a fit from
turbulence dynamics although being able to reproduce experimental
data to more or less accuracy. A deductive support from the
three-dimensional Navier-Stokes equation was stressed to be
essential for the fitting procedure to be considered meaningful.

Recently Laval, Dubrulle, and Nazarenko~\cite{Laval} have
developed a stochastic kind of Batchelor-Proudman rapid distortion
theory approach to the three-dimensional Navier-Stokes equation
using separation of large-scale and small-scale velocities and
Gabor transformation (localized wave-packets) to derive
one-dimensional Langevin toy model for small-scale velocity
increments both in the equally interesting Eulerian and Lagrangian
frames; see also recent paper~\cite{Laval2}. The large-scale terms
entering the resulting approximate small-scale equation are
related to large-scale strain and inter-scale coupling and are
treated as noises with a given statistics. The small scales are
stochastically distorted in certain way as a combined effect of
the large scales and the inter-scale coupling. Short-time
correlated character of the distortions follows from the numerical
study of decaying turbulence; longtime correlations and dependence
on velocities which may be present here as well have not been
modeled in the first step. This approach allows one to account for
nonlocal interaction effects in small-scale turbulence via a
simple random multiplicative process driven by coupled Gaussian
white-in-time multiplicative and additive noises, while local
small-scale interactions are modeled by a turbulent viscosity.

In a comparative analysis of some recent one-dimensional Langevin
toy models of fluid particle acceleration in developed
three-dimensional turbulence~\cite{Aringazin5} we have
demonstrated that the one-dimensional Laval-Dubrulle-Nazarenko
(LDN) type model~\cite{Laval,Laval2}, with the model turbulent
viscosity $\nu_{\mathrm t}$ and delta-correlated Gaussian white
multiplicative and additive noises, formulated for the Lagrangian
acceleration meets the experimental data on acceleration
statistics~\cite{Bodenschatz,Bodenschatz2} to a good accuracy.
Particularly, it was shown that the resulting contribution to
fourth order moment, $a^4P(a)$, {does} peak at the same values as
the experimental curve, in contrast to predictions of the most of
other stochastic models~\cite{Beck,Beck4,Aringazin3,Reynolds}.

Also, within the framework of Random Intensity of Noise (RIN)
approach~\cite{Aringazin5} to the LDN type model the assumption
that the additive noise intensity $\alpha$ depends on absolute
value of velocity fluctuations $u$ was found to imply the
conditional probability density function $P(a|u)$ and the
conditional acceleration variance $\langle a^2|u\rangle$ which are
in a good qualitative agreement with the recent experimental data
on the conditional acceleration statistics reported by Mordant,
Crawford, and Bodenschatz~\cite{Mordant0303003}. These results
have been obtained in the particular case when the correlation
between the multiplicative and additive noises is taken to be
zero.

In the present paper, we fill the gap by studying the effect of
nonzero cross correlation of the noises that models a relationship
between stretching and vorticity in the three-dimensional
case~\cite{Laval} and can be seen as a skewness of the probability
density function of acceleration component.

We remind that shell models of turbulence are incapable to
describe the observed skewness generation along the scale of the
probability density function of the longitudinal Eulerian velocity
increments~\cite{Laval}. This skewness is of much importance in
understanding of the Richardson-Kolmogorov cascade picture of the
homogeneous isotropic turbulence. The skewness is particularly
related to a non-zero value of the Eulerian third-order velocity
structure function, and therefore to the essence of the turbulent
cascade via the Kolmogorov four-fifths
law~\cite{Laval,MarcqPF2001,MoisyPRL1999}.

Intermittency of homogeneous and isotropic turbulence is usually
characterized by a nonlinear dependence of the scaling exponents
on the order $n$ of moment, and is well established experimentally
both in the Eulerian and Lagrangian frameworks for approximately
homogeneous and isotropic flows.

The Eulerian and Lagrangian anomalous scalings trace back to a
local inhomogeneity of the flow and longtime correlations in the
particle accelerations respectively~\cite{Mordant0206013}. The
probability density functions of the Eulerian and Lagrangian
velocity increments both exhibit the same behavior: they are
approximately Gaussian at large scales and progressively develop
stretched exponential type tails when the spatial and time
increments decrease down to the Kolmogorov length and time,
respectively. In the limit of zero time increments the probability
density function of the Lagrangian velocity increments converges
to that of the Lagrangian acceleration. In practice, one uses
nonzero time increments $\tau$ lying within the range of strong
viscous dissipation, such that velocities are smoothed and the
characteristic relation $u(t+\tau)-u(t)=\tau a(t)$ holds to a good
accuracy. This time scale is known to be about or less than the
Kolmogorov time.

In the Eulerian framework, the relative scaling exponents of the
absolute moments of the longitudinal Eulerian velocity increments,
$\langle|u(x+r)-u(x)|^n\rangle$, were measured to be $\zeta_n^E$ =
0.36, 0.70, 1.28, 1.53 for $n=1,2,4,5$~\cite{Mordant0103084}, and
the third-order Eulerian velocity structure function, taken as a
reference, is known to scale linearly with the spatial separation
$r$ due to the Kolmogorov scaling; see, however,
Ref.~\cite{MoisyPRL1999}, in which Kolmogorov-Novikov equation
accounting for the external forcing length in the $r^2$ term has
been confirmed experimentally for an impressive range of Reynolds
numbers to a very high accuracy.

This picture converts into the Lagrangian framework.
The relative scaling exponents of the absolute moments of the
Lagrangian velocity increments (one component),
$\langle|u(t+\tau)-u(t)|^n\rangle$, were measured to be
$\zeta_n^L$ = 0.56$\pm$0.01, 1.34$\pm$0.02, 1.56$\pm$0.06,
1.8$\pm$0.2 for $n=1,3,4,5$~\cite{Mordant0103084}, and the
second-order Lagrangian velocity structure function, taken as a
reference, is known to scale linearly with the time increment
$\tau$ due to the dimensional analysis. One therefore expects
skewness of the probability density function of the Lagrangian
velocity increments in time since $\zeta_3^L$ is nonzero, and
hence that of the Lagrangian acceleration.

The one-dimensional LDN toy model was formulated originally in
both the Eulerian and Lagrangian frameworks for velocity
increments in the frame comoving with the wavepacket~\cite{Laval}.
We use the Lagrangian formulation and the exact result for
probability density function for LDN type model obtained as a
stationary solution of the Fokker-Planck equation associated to
the one-dimensional Langevin equation for the acceleration
component~\cite{Laval,Aringazin5},
\be\label{LangevinLaval}
\partial_t a = (\xi - \nu_{\mathrm t}k^2)a + \sigma_\perp.
\ee
Here, the Gaussian white noises $\xi$ and $\sigma_\perp$ model
stochastic forces in the Lagrangian frame and are defined by
\begin{eqnarray}\label{noises}
\langle\xi(t)\rangle=0, \
\langle\xi(t)\xi(t')\rangle = 2D\delta(t-t'), \nonumber \\
\langle\sigma_\perp(t)\rangle = 0, \
\langle\sigma_\perp(t)\sigma_\perp(t')\rangle = 2\alpha\delta(t-t'), \\
\langle\xi(t)\sigma_\perp(t')\rangle = 2\lambda\delta(t-t'). \nonumber
\end{eqnarray}
All the noises are treated along a particle trajectory. Here, the
free parameters $\alpha$, $D$, and $\lambda$ measure intensities
of the noises and their cross correlation, respectively.

Such a choice of noises is motivated not only by simplicity of
their statistics but also by the DNS of decaying
turbulence~\cite{Laval}.

Similar problem within the Eulerian framework has been recently
investigated, with the result that the noise entering Langevin
type equation can be safely taken delta-correlated in scale.
Namely, the Eulerian experimental study of Langevin modeling of
velocity increments by Renner, Friedrich and
Peinke~\cite{Friedrich} and Marcq and Naert~\cite{MarcqPF2001}
reveals that the longitudinal velocity is correlated over
distances much larger than the correlation length of its spatial
derivative, so that the Markovian approximation is accurate in the
inertial range. Approximation of a short-correlated noise by the
delta-correlated one is usually made due to the scale hierarchy
validating the use of Langevin type equations. It is natural to
map this result to the Lagrangian domain.

In the Eulerian LDN framework, the cross correlation parameter was
found to control third-order longitudinal velocity structure
function due to a kind of generalized Karman-Hovarth relationship
derived in Ref.~\cite{Laval}. Since the noise distributions are
taken to be not skewed, this suggests that the parameter $\lambda$
defined in Eq.~(\ref{noises}) should be nonzero.

Experiments~\cite{MarcqPF2001} show that the distribution of the
Eulerian longitudinal velocity increments is slightly asymmetric
in the inertial range of scales, with the skewness factor being
about $S=-0.25$ for the studied $R_\lambda=430$ flow; the
odd-order moments ($2n+1$) are known to be small as compared with
even-order ones ($2n+2$). The Langevin model of
Ref.~\cite{MarcqPF2001} assumes the use of only one noise driving
Eulerian velocity increments across scales, and this noise is
characterized by a slightly skewed distribution (the skewness
factor $S=0.55$) and long tails (the flatness factor $F=8.5$ as
compared to $F=3$ for a Gaussian). Gaussian approximation for this
noise was used to derive the Fokker-Planck equation. It was found
that this approximation and accounting for corrections coming from
non-Gaussianity of the noise imply some persistent deviations from
the observed scaling of third-order and fifth-order velocity
structure functions. This may indicate that some different type of
Langevin or Fokker-Planck equation should be used as ansatz.

The probability density function as a stationary solution of the
Fokker-Planck equation associated to Eqs.~(\ref{LangevinLaval})
and (\ref{noises}) was calculated exactly~\cite{Aringazin5},
\be\label{PLaval} P(a) = \frac{C \exp[-{\nu_{\mathrm
t}k^2}/{D}+F(c)+F(-c)]} {(Da^2\!-\!2\lambda a
\!+\!\alpha)^{1/2}(2Bka+\nu_{\mathrm t}k^2)^{{2B\lambda
k}/{D^2}}},
\ee
where we have denoted
\begin{eqnarray}\label{A4}
F(c)
 = \frac{c_1k^2}{2c_2D^2c}\ln[\frac{2D^3}{c_1c_2(c-Da+\lambda)}
   \nonumber \\
   \times(
   B^2(\lambda^2 + c\lambda-D\alpha)a
   + c(D\nu_{\mathrm t}^2k^2+c_2\nu_{\mathrm t})
    )
   ],\\
c=-i\sqrt{D\alpha-\lambda^2}, \quad \nu_{\mathrm t} =
\sqrt{\nu_0^2+ B^2a^2/k^2},\\
c_1 = B^2(4\lambda^3\!+\!4c\lambda^2\! -\!
3D\alpha\lambda-cD\alpha)
    \!+\! D^2(c\!+\!\lambda)\nu_0^2k^2,\\
c_2 = \sqrt{B^2(2\lambda^2 + 2c\lambda-D\alpha)k^2 +
D^2\nu_0^2k^4},
\end{eqnarray}
and $C$ is normalization constant.

Without loss of generality one can put $k=1$ and $\alpha=1$ by
appropriate rescaling of the parameters $D$, $B$, $\nu_0$, and
$\lambda$~\cite{Aringazin5} to make a fit of the above $P(a)$ to
the experimental data. As one can see from Eq.~(\ref{PLaval}), the
parameter $\lambda$ introduced in Eq.~(\ref{noises}) is
responsible for an asymmetry of the distribution with respect to
$a\to-a$.

The fit for the particular (symmetric) case, $\lambda=0$, has been
made in Ref.~\cite{Aringazin5}. The flatness factor of the studied
distribution which characterizes the widening of its tails (when
compared with a Gaussian) is found to be $F=42.5$ 
(for $k = 1$, $\alpha=1$, $D = 1.130$, $B = 0.163$, $\nu_0 =
2.631$, $C=1.805$) that deviates from the flatness of the
experimental curve, $F=55\pm8$~\cite{Mordant0303003}. In numerical
calculations we used a cutoff by restricting the integration range
by $|a|/\langle a^2 \rangle^{1/2}\leq 1000$.

The RIN approach extends the LDN type model (\ref{LangevinLaval})
by assuming certain relationship of noise intensities and in
general other model parameters to velocity fluctuations $u$, and
enables one to study acceleration statistics conditional on
velocity fluctuations~\cite{Aringazin5}. A dependence of
acceleration distribution on velocity fluctuations is known to
violate Kolmogorov 1941 local homogeneity of the
flow~\cite{Mordant0303003}.

The paper is organized as follows.

In Sec.~\ref{Sec:Skewness} we study a skewness effect implied by a
nonzero cross correlation parameter $\lambda$, both for
unconditional and conditional probability density functions of the
acceleration component. In Sec.~\ref{Sec:ConditionalMean} we study
conditional mean acceleration using the RIN extension of the LDN
type model (\ref{PLaval}). In Sec.~\ref{Sec:Conclusions} we
summarize the obtained results and make a few concluding remarks.

\section{The skewness}\label{Sec:Skewness}

\subsection{The unconditional probability density function}
\label{Sec:Unconditional}

\begin{figure}[tbp!]
\begin{center}
\includegraphics[width=0.45\textwidth]{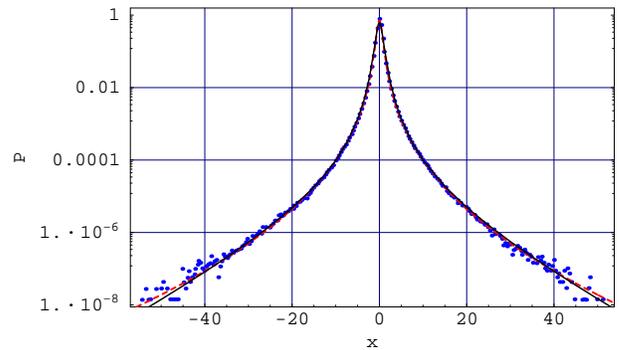}
\end{center}
\caption{ \label{Fig1} Acceleration probability density function
$P(a)$. Dots: the experimental data for the transverse component
of acceleration at $R_\lambda=690$ by Crawford, Mordant, and
Bodenschatz~\cite{Bodenschatz2}.  Dashed line: stretched
exponential fit~\cite{Bodenschatz2}, $\beta=0.513$, $\sigma=
0.563$, $\gamma= 1.600$, $C=0.733$. Solid line: The model
(\ref{PLaval}), $k=1$, $\alpha=1$, $D=1.100$, $B=0.155$,
$\nu_0=2.910$, $\lambda=-0.005$, $C=3.230$. $x=a/\langle a^2
\rangle^{1/2}$ denotes normalized acceleration.}
\end{figure}

\begin{figure}[tbp!]
\begin{center}
\includegraphics[width=0.45\textwidth]{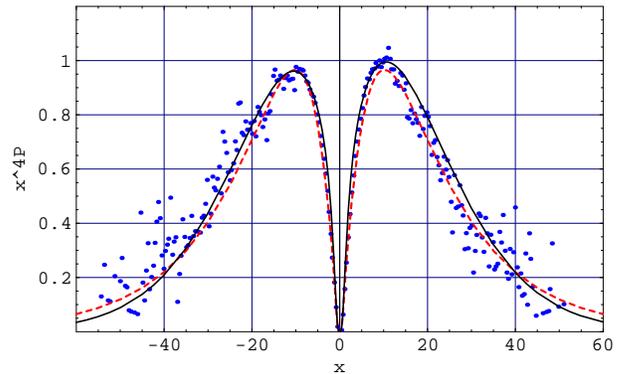}
\end{center}
\caption{ \label{Fig2} The contribution to fourth order moment,
$a^4P(a)$. Notation is the same as in Fig.~\ref{Fig1}.}
\end{figure}

In the present paper we generalize the consideration made in
Ref.~\cite{Aringazin5} by letting the cross correlation parameter
to be nonzero, $\lambda\not=0$. For this case, the exact
probability density function~(\ref{PLaval}) will be used.

A sample fit of the distribution (\ref{PLaval}) is presented in
Fig.~\ref{Fig1} and the contribution to the fourth order moment is
plotted in Fig.~\ref{Fig2}, in which the signature of a skewness
of the experimental distribution (dots) of the acceleration
component can be seen as a small difference of about 5\% in
heights of the two peaks. Geometrically, this component of
acceleration corresponds to the direction transverse to the axial
symmetry ($z$) axis of the large-scale forcing of studied
flow~\cite{Bodenschatz}.

We remind that the turbulence was generated in a flow between
counter-rotating disks in a cylindrical container, and the flow
significantly deviates from the ideal of homogeneity and isotropy.
By the large-scale flow symmetry the two transverse components
($x$ and $y$) are taken statistically equivalent, and distinct
from the axial component ($z$), in the Cartesian laboratory frame
of reference. Particle accelerations were measured in a small
volume in the center of the flow chamber within which the flow is
approximately homogeneous and isotropic. Only one transverse ($x$)
and the axial ($z$) components were actually measured. Statistical
properties of the unmeasured $y$ component are expected to be
identical to those of the $x$
component~\cite{Bodenschatz,Bodenschatz2}. Throughout the paper we
consider the data for the $x$ component of acceleration.

One observes a better agreement of the skewed curve (solid line in
Fig.~\ref{Fig2}) for the fitted value
\be\label{lambdafit}
\lambda=-0.005
\ee
($k=1$, $\alpha=1$, $D=1.100$, $B=0.155$, $\nu_0=2.910$,
$C=3.230$) with the data points in the intermediate range of
positive and negative accelerations, $|a|/\langle a^2\rangle^{1/2}
\simeq 10$, as compared with (symmetric) curves implied by the
stretched exponential fit (dashed line in
Fig.~\ref{Fig2})~\cite{Bodenschatz2}, the RIN chi-square Gaussian
model fit~\cite{Aringazin3}, the RIN log-normal
model~\cite{Beck3}, and the Reynolds model~\cite{Reynolds}.

We note that better result was recently
obtained~\cite{Bodenschatz2} by
processing the original (unfitted) curve of Ref.~\cite{Reynolds}
in exactly the same way as that for the data but some departure
from the experimental data still persists. Particularly, the
associated contribution to fourth order moment, $a^4P(a)$, does
not peak at the same value as that for the experimental curve,
being however very close to it.

Since in statistically homogeneous and isotropic turbulent flows
for the $x$, $y$, or $z$ component of acceleration one should have
zero skewness, we could attribute the observed small skewness to
anisotropy of the studied $R_{\lambda}=690$ flow.

We remind that the relative root-mean-square ({\it rms})
uncertainty of the experimental $P(a)$ is about 3\% for
$|a|/\langle a^2\rangle^{1/2} \leq 10$ (the most accurate part of
the distribution), and is less than 40\% for $|a|/\langle
a^2\rangle^{1/2} \leq 40$~\cite{Mordant0303003}. Clearly, less
uncertainties are required to obtain good quantitative description
of the skewness as it appears to be a quite tiny effect.
Nevertheless, in the present section we use the available
experimental data to {\em demonstrate} the effect of nonzero
$\lambda$.

Recently reported high precision data~\cite{Mordant0303003} show
that the mean acceleration conditional on velocity fluctuations is
nonzero and increases with the increase of velocity. This was
claimed to be related to the anisotropy of the studied flow,
although it was pointed out that DNS of homogeneous isotropic
turbulence also shows slightly nonzero mean acceleration. We will
consider this issue below and in Sec.~\ref{Sec:ConditionalMean}.


It should be emphasized that for the studied $R_{\lambda}=690$
flow the ratio between the variances of the $x$ and $z$ components
of acceleration was measured to be slightly different from unity
due to $a_{0x}/a_{0z}\simeq 1.06$~\cite{Bodenschatz}. It was
pointed out that experimental biases do not affect acceleration
measurements (they depend mostly on the velocity) except to the
extent that the acceleration and velocity are correlated. The
shapes of the probability distributions for the $x$ and $z$
components of acceleration were measured to be approximately the
same at high Reynolds numbers. This level of $x$-to-$z$ anisotropy
was found to persist for higher Reynolds numbers (data presented
for $R_{\lambda}=970$) that may indicate a fundamental character
of this phenomenon. Namely, while the K41 theory postulates
complete universality it is still an open question to what extent
statistical properties of the 3D turbulence in the inertial range
do not depend on the details of large-scale forcing. As the
presence of finite injection scale is felt through the entire
inertial range via the anomalies of the Eulerian and Lagrangian
scaling exponents (K62) it is natural to expect that the
statistics of fluid particle acceleration, which is generally
associated to small scales of the flow, should reflect the
anisotropy of the large scale forcing. From this point of view,
the observation supports the view that the induced anisotropy in
the acceleration statistics is a rule rather than exception in the
context of developed turbulence dynamics: Forced anisotropy at
large scales is not washed out in the inertial range of a
high-Reynolds-number flow and seems to be felt at small scales,
smaller than the Kolmogorov length. This kind of anisotropy in
acceleration statistics may not be directly related to the
parameter $\lambda$.

The cross correlation parameter $\lambda$ is nonzero for
longitudinal and zero for transverse velocity increments by
construction~\cite{Laval}. For the Lagrangian acceleration, this
reads that for the component of Lagrangian acceleration, $a_\tau$,
pointed along the corresponding Lagrangian velocity at some point
of the particle trajectory the parameter $\lambda$ is nonzero
while for the transverse component $a_n$ it is zero. The
experimental data on $a_x$ and $a_z$ time series do not allow one
to extract separately $a_\tau$ and $a_n$, in order to verify that
the $a_\tau$ acceleration PDF is skewed while the $a_n$
acceleration PDF is symmetric relative to the change of sign of
acceleration. This requires obtaining the data on the whole set of
the acceleration and corresponding velocity components, to have an
access to geometry of individual trajectories of the tracer
particle. Also, one should suppress the effect of above mentioned
induced anisotropy. The influence of the large-scale anisotropy on
$a_\tau$ and $a_n$ acceleration PDFs may occur to be different. In
contrast to the experiment where it seems to be difficult to
provide sufficiently high level of isotropy and extract
statistical data on $a_\tau$ and $a_n$, the DNS could be used to
test skewness of the $a_\tau$ and $a_n$ acceleration PDFs. The
isotropy is well satisfied in DNS and at least it is free from
strong large-scale anisotropy of the experimentally studied von
Karman flow.

The obtained small value (\ref{lambdafit}) of the fitted cross
correlation parameter $\lambda$, as compared with the used values
of the noise intensities $D$ and $\alpha$, is in a good agreement
with the results of numerical Rapid Distortion Theory (RDT)
analysis of the noise cross correlators in decaying
turbulence~\cite{Laval}. Particularly, $\lambda$ turned out to be
about two orders of magnitude smaller than $D$ and $\alpha$. The
time scale of the cross correlation as well as of the
autocorrelation of additive noise.

Below we use the RIN extension~\cite{Aringazin5} of the LDN model
(\ref{LangevinLaval}) to obtain and study the conditional
probability density function.

\subsection{The conditional probability density function}
\label{Sec:Conditional}

In the RIN approach the result (\ref{PLaval}) is treated in
general as a probability density function {\em conditional} on the
parameters involved in the model. It was found~\cite{Aringazin5}
that for $\lambda=0$ only a variation of the additive noise
intensity $\alpha$, with $\alpha=e^{u/u_0}$, qualitatively meets
(i) the observed variation of the shape of experimental
probability density function of the transverse component of
acceleration conditional on the transverse component of velocity
fluctuations $u$ with variation of $u$, and (ii) the polynomial
type increase of the normalized conditional acceleration variance
with an increase of $|u|$~\cite{Mordant0303003}.

In the present paper we study the effect of nonzero $\lambda$. In
general one may expect that the variation of $\lambda$ causes not
only essential variation of the skewness but at the same time a
considerable variation of the acceleration variance.

\begin{figure}[tbp!]
\begin{center}
\includegraphics[width=0.45\textwidth]{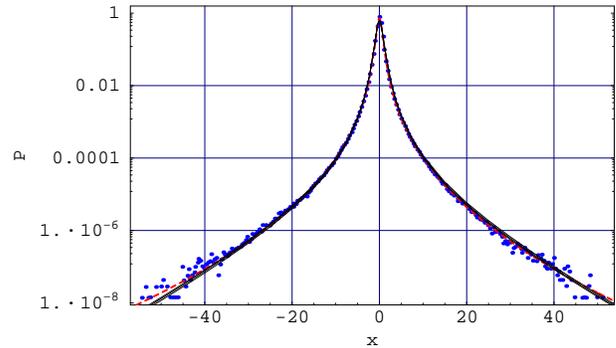}
\end{center}
\caption{ \label{Fig3} The acceleration probability density
function. Dots: the experimental data for the transverse component
of acceleration at $R_\lambda=690$ by Crawford, Mordant, and
Bodenschatz~\cite{Bodenschatz2}.  Dashed line: stretched
exponential fit~\cite{Bodenschatz2}, $\beta=0.513$, $\sigma=
0.563$, $\gamma= 1.600$, $C=0.733$. Solid lines: the model
(\ref{PLaval}) at $\lambda=-0.005, -0.025, -0.05$ ($k=1$,
$\alpha=1$, $D=1.100$, $B=0.155$, $\nu_0=2.910$). $x=a/\langle a^2
\rangle^{1/2}$.}
\end{figure}

\begin{figure}[tbp!]
\begin{center}
\includegraphics[width=0.45\textwidth]{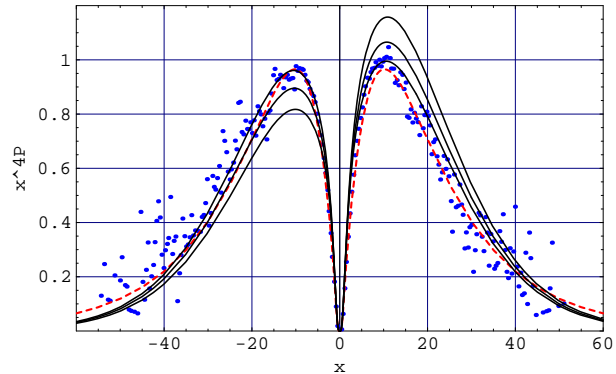}
\end{center}
\caption{ \label{Fig4} The contribution to fourth order moment,
$a^4P(a)$. Notation is the same as in Fig.~\ref{Fig3}.}
\end{figure}

The experimental conditional distributions $P(a|u)$ for $u$
ranging from 0 to 3 {\it rms} velocity were found to be almost of
the same shape as the experimental unconditional distribution
$P(a)$~\cite{Mordant0303003}. This means that they share the same
process underlying Lagrangian turbulence intermittency.

In a qualitative analysis, we can use the fact that the
experimental unconditional distribution $P(a)$ is approximately of
the same form as the experimental conditional distribution
$P(a|u)$ at $u=0$. As we suppose $\lambda=\lambda(u)$ the net
effect of a variation of $u$ will be some variation of the
parameter $\lambda$ in $P(a)$. For this case, the result of our
check is that the variation of the shape of curve with variation
of $\lambda$ around the value $\lambda=-0.005$ (the other
parameters fixed) is not in a qualitative agreement with the
experimental data on $P(a|u)$ shown in Fig.~6a of
Ref.~\cite{Mordant0303003}. Namely, the obtained plots shown in
Figs.~\ref{Fig3} and \ref{Fig4} demonstrate that the variation of
$\lambda$ affects mainly the skewness rather than the variance of
the distribution (\ref{PLaval}).

We conclude that within the framework of RIN approach the cross
correlation parameter $\lambda$ (as well as the parameters $D$,
$\nu_0$, and $B$ as shown in Ref.~\cite{Aringazin5}), with
$\lambda=\lambda(u)$, could not be responsible for the specific
visible change of the shape of experimental conditional
acceleration probability density function with an increase of
velocity fluctuations~\cite{Mordant0303003}.

We are thus left with the only possibility: to assign the observed
{\em essential} dependence of $P(a|u)$ on velocity fluctuations
$u$ to the additive noise intensity $\alpha$. In general, this is
in an agreement with the RDT approach by Laval, Dubrulle, and
Nazarenko, particularly with the approximate LDN model
(\ref{PLaval}), in which only the additive noise intensity is
characterized by a dependence on small scale velocity fluctuations
coupled to large scale velocity fluctuations.

It should be emphasized here that in the original LDN model the
other parameters of the model, $B$, $\nu_0$, and $\lambda$, were
not assumed to depend on small scale velocity fluctuations, and
the multiplicative noise intensity $D$ was found to depend only on
large scale velocity fluctuations. Nevertheless, in
Ref.~\cite{Aringazin5} we have evaluated the effect caused by each
of these parameters, $D(u)$, $B(u)$, $\nu_0(u)$, and $\lambda(u)$,
in order to verify independently whether this may qualitatively
correspond to the set of experimental conditional distributions
$P(a|u)$, $u/\langle u^2\rangle^{1/2} \in [0,3]$. Despite we
obtained a negative result, presence of some weak dependencies of
these parameters on $u$ can not be ruled out on the basis of made
qualitative comparison.

Adopting the point of view that $\alpha=\alpha(u)$ as a first
approximation, in the next section we turn to a consideration of
the mean acceleration conditional on velocity fluctuations $u$.

\section{The conditional mean acceleration}\label{Sec:ConditionalMean}

The conditional mean acceleration should be zero in homogeneous
and isotropic turbulence, and departures from zero reflect the
anisotropy of the studied flow although DNS of homogeneous
isotropic turbulence has also shown slight departures from
zero~\cite{Mordant0303003}.

As the first step, the conditional mean acceleration can be
calculated under the assumption that only $\alpha$ depends on
velocity fluctuations $u$. We take an exponential dependence,
$\alpha=e^{u/u_0}$, $u_0=3$, which was found to be relevant from
both the theoretical and experimental points of
view~\cite{Aringazin5}.

\begin{figure}[tbp!]
\begin{center}
\includegraphics[width=0.45\textwidth]{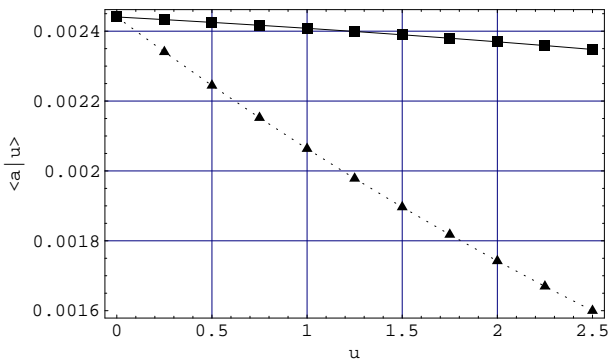}
\end{center}
\caption{ \label{Fig5} The conditional mean transverse
acceleration as a function of standardized velocity fluctuations
$u$ for $\alpha=e^{|u|/3}$; $k=1$, $D=1.100$, $B=0.155$,
$\nu_0=2.910$, $\lambda=-0.005$. Triangles: normalized to unit
conditional acceleration variance, $\langle a|u\rangle/\langle
a^2|u\rangle^{1/2}$. Boxes: normalized to unit conditional
acceleration variance at $u=0$, $\langle a|u\rangle/\langle
a^2|0\rangle^{1/2}$.}
\end{figure}

Assuming that the unconditional distribution $P(a)$ is
approximately of the same shape as the conditional distribution
$P(a|u)$ at $u=0$, and using the same set of fitted parameters as
in Sec.~\ref{Sec:Unconditional} ($\lambda=-0.005$) we obtain that
the conditional mean acceleration only slightly deviates from
zero,
\be
\frac{\langle a|u\rangle}{\langle a^2|u\rangle^{1/2}} \simeq
0.002,
\ee
for all values of $|u|$ ranging from zero to 2.5 {\it rms}
velocity, with the tendency to {\em decrease} down to zero with
the increase of $|u|$, as shown in Fig.~\ref{Fig5} (triangles).

In general, this is in agreement with almost symmetric shapes of
the experimental probability density functions of the component of
acceleration conditional on the same component of velocity
fluctuations~\cite{Mordant0303003}. Alas, an illustrative
character of the presented experimental plots and the increasing
experimental uncertainty of $P(a|u)$ at big $|u|/\langle
u^2\rangle^{1/2}$ does not allow us to make a definite conclusion
since the skewness effect is very small to be readily seen from
the experimental $P(a|u)$. Also, we note that the predicted value
$\langle a|0\rangle/\langle a^2|0\rangle^{1/2} \simeq 0.0024$ (see
Fig.~\ref{Fig5}) does not contradict to that shown in Fig.~6b of
Ref.~\cite{Mordant0303003}, $\langle a|0\rangle/\langle
a^2\rangle^{1/2} \leq 0.05$ (uncertainty of the presented data
points does not allow us to give the upper bound more precisely).

The above result on the conditional mean transverse acceleration
(Fig.~\ref{Fig5}) is however in a sharp contrast with the reported
experimental dependence of the conditional mean transverse
acceleration on $u$ (Fig.~6b of Ref.~\cite{Mordant0303003}) which
displays that $\langle a|u\rangle/\langle a^2\rangle^{1/2}$ {\em
increases} from about zero at $|u|=0$ to about 0.3 at $|u|/\langle
u^2\rangle^{1/2}=2.5$, in some nonlinear way.

In essence the experimental data show that with the increase of
velocity fluctuations $|u|$ the mean of the conditional transverse
acceleration distribution $P(a|u)$ normalized to unit
unconditional acceleration variance increases by small but
appreciable amount. One would like to know whether this holds for
the $z$ component of acceleration. We expect that the mean for
this component is bigger than that for the $x$ component,
partially due to about 3\% smaller variance of the $z$ component
of acceleration~\cite{Bodenschatz}.

More detailed study is required to explain this highly remarkable
phenomenon, which may indicate stronger coupling of the
multiplicative noise to the additive one for bigger velocity
fluctuations $|u|$ in the Lagrangian frame.

Below, we use definitions of the noises provided by the LDN
approach~\cite{Laval} and the results for the velocity-dependence
of noise intensities~\cite{Aringazin5} to give a tentative
explanation of this phenomenon.

The additive noise intensity considerably increases for bigger
$|u|$ due to the relationship $\alpha \sim e^{|u|}$ (incoherent
noisy background responsible for the random walk behavior of
acceleration is much intensified) while the multiplicative noise
intensity remains at approximately the same level (very intense
vortical structures responsible for the random multiplicative
process are relatively frozen in time and characterized by a
saturation level of $|u|$ for a given Reynolds number and
vorticity). This means an increase of large scale effects produced
by the interaction between small and large scales (the effect of
nonlocal interactions). Since the large scales are the only
unifying agent between the noises the cross correlation of the
noises becomes more pronounced for bigger $|u|$. Whereas a direct
effect of the increase of additive noise intensity tends to
symmetrize the distribution (see Fig.~\ref{Fig5}) due to a higher
degree of chaoticity (higher statistical isotropy), bigger $|u|$
may imply a variation of the cross correlation, to which the
skewness is highly sensitive, so that the overall effect is a
small but appreciable increase of the mean of the conditional
distribution. To be more precise, the net effect depends on
competition between $\lambda$ and $\alpha$ as the velocity $|u|$
increases.

Also, it is worthwhile to note that the effect of discrete
Kolmogorov turbulent cascade may be of importance here since it is
characterized by a relationship between high-amplitude harmonics
of the basic ratio (nonlocal interactions).

The following remark is in order.

Note that in Sec.~\ref{Sec:Conditional} we used the experimental
data on unconditional $P(a)$ instead of that for the conditional
$P(a|u)$ with the aim to estimate result of the possible
dependence of $\lambda$ on $u$. Such an approach is fully
justified since these curves are of a similar character and we
were interested only in the general effect of the variation of
$\lambda=\lambda(u)$ on the shape of $P(a|u)$, for which case
particular values of the fitted parameters are not important. This
dependence was found to be not capable to explain the
characteristic variation of the shape of $P(a|u)$ with variation
of $u$ shown in Fig.~6a of Ref.~\cite{Mordant0303003}. In other
words, it does not make considerable contribution to the variance
of $P(a|u)$ consistent with the experimental data. However, this
evidently does not exclude possible presence of a specific, {\it
e.g.}, polynomial or exponential, dependence of $\lambda$ on $u$
which would allow one to qualitatively explain the observed small
variation of the conditional mean acceleration $\langle
a|u\rangle$ with variation of $u$.

Note that we have evaluated the conditional mean acceleration
normalized to unit conditional acceleration variance $\langle
a^2|u\rangle^{1/2}$ rather than to unit unconditional acceleration
variance $\langle a^2\rangle^{1/2}$. This makes a difference since
the conditional acceleration variance $\langle a^2|u\rangle^{1/2}$
increases with an increase of
$u$~\cite{Mordant0303003,Aringazin5}, contrary to the case of
unconditional acceleration variance that is constant.

In the RIN approach, the evaluation of unconditional acceleration
variance requires a calculation of the stochastic expectation of
$P(a|u)$ over random $u$. Whereas we can judiciously choose, in
the first approximation, the distribution of $u$ to be Gaussian
with zero mean and $\alpha=e^{u/u_0}$~\cite{Aringazin4,Aringazin5}
we should also take into account for the emerging effect of
$\lambda$, for which case we have no clue to make a judicious
choice of some function $\lambda(u)$. An accounting for
$\lambda(u)$ may produce a considerable contribution to $P(a)$ due
to the integral effect. For this reason, in the present paper we
do not evaluate $P(a)$ unless we identify the functional form of
$\lambda(u)$. This issue is of much interest and can be studied
elsewhere.

Nevertheless, one can proceed here with a model-independent study
by normalizing the conditional mean acceleration to unit
conditional acceleration variance at $u=0$ which is {\em constant}
(does not depend on $u$), {\it i.e.}, by evaluating $\langle
a|u\rangle/\langle a^2|0\rangle^{1/2}$. The result is plotted in
Fig.~\ref{Fig5} (boxes) and shows a much steeper decrease as
compared with that of $\langle a|u\rangle/\langle
a^2|u\rangle^{1/2}$ (triangles), but yet {no increase}.

In summary, the predicted conditional mean acceleration $\langle
a|u\rangle/\langle a^2|0\rangle^{1/2}$ is negligibly small and
decreases down to zero for larger $|u|$ under the assumption that
$\alpha=e^{|u|/3}$ and the other parameters to be constant. Hence,
the additive noise tends to symmetrize acceleration distribution.

Therefore, we are left with the only possibility to explain the
observed appreciable increase of $\langle a|u\rangle/\langle
a^2\rangle^{1/2}$ with the increase of $|u|$: to assign some
dependence of the parameter $\lambda$, which measures correlation
between the additive noise and the multiplicative one, on velocity
fluctuations $u$. It should be noted that this is in agreement
with the LDN model, in which the additive noise depends on
small-scale velocity fluctuations.

It should be stressed however that as it has been mentioned above
the observed nonzero mean acceleration is associated to the flow
anisotropy which may be not related to the effect described by the
parameter $\lambda$.

\section{Discussion and conclusions}\label{Sec:Conclusions}

(i) We have shown that the cross correlation parameter $\lambda$
of the LDN type model (\ref{PLaval}) for a particle acceleration
could be used to explain a skewness of the acceleration
probability density function, and estimated its value,
$\lambda\simeq -0.005$, by using a fit to the recent experimental
statistics data on the transverse component of acceleration.

(ii) The mean acceleration is found to be very close to zero,
$\langle a|u\rangle/\langle a^2|0\rangle^{1/2} \leq 0.0024$, when
the predicted $a^4P(a)$ is fitted to the experimental data. The
mean acceleration should vanish for homogeneous isotropic
turbulence. The observed mean acceleration can be attributed to
small anisotropy (imperfection) of the studied $R_\lambda=690$
flow. However, the DNS, for which isotropy is well satisfied,
indicates slight departures from zero. Whether this is of some
importance for developed turbulence should be clarified. In any
case, the observed noticeable nonzero mean of Lagrangian
acceleration, which is usually associated to extremely small
scales of the flow, in the high-Reynolds-number flow that is
anisotropic at large scale, deserves a separate study.

(iii) Using the RIN approach which extends the LDN type model by
assuming certain relationship of noise intensities and in general
other model parameters to velocity fluctuations $u$ we have
studied acceleration statistics conditional on velocity
fluctuations. We found that the assumption $\lambda=\lambda(u)$
could not be responsible for the experimentally observed
characteristic variation of the shape of conditional distribution
$P(a|u)$ with variation of $u$. Taken together with the result of
our previous work~\cite{Aringazin5} this implies that only the
additive noise intensity $\alpha$ reveals an essential dependence
on $u$ for which we used the exponential function, $\alpha =
e^{u/u_0}$, relevant from both the phenomenological and
experimental points of view. The additive noise tends to
symmetrize acceleration distribution for larger $|u|$: the
simulated conditional mean acceleration is very small and
decreases down to zero as shown in Fig.~\ref{Fig5}. This is not in
agreement with the experimental data even qualitatively. However,
the observed mean acceleration could be attributed mainly to
anisotropy (imperfection) of the studied $R_\lambda=690$ flow.

(iv) We have found that despite the previously obtained
result~\cite{Aringazin5} that a sole dependence of the additive
noise intensity parameter $\alpha$ on $u$ is in a good qualitative
agreement with the experimental data on the conditional
acceleration variance, $\langle a^2|u\rangle/\langle a^2\rangle$,
it appears to be not capable to capture qualitatively the
experimental data on the conditional mean acceleration, $\langle
a|u\rangle/\langle a^2\rangle^{1/2}$, which increases for bigger
$|u|$. While determination of the conditional mean acceleration
for higher isotropic turbulent flows can be left for future
experiments and numerical simulations, we speculate on the
relationship between $\lambda$ and skewness. This suggests that,
in addition to the dependence $\alpha=\alpha(u)$, the cross
correlation parameter $\lambda$, which directly controls the
skewness, could depend on $u$ in some way.

(v) As the result, to meet the available experimental data and the
DNS we are led to consider, in a self-consistent RIN approach, the
conditional probability density function (\ref{PLaval}) in the
form $P(a|u)=P(a|\alpha(u),\lambda(u))$, with $\alpha=e^{u/u_0}$,
velocity fluctuations $u$ to be Gaussian distributed with zero
mean, some function $\lambda(u)$ to be determined, and free
parameters $D$, $B$, and $\nu_0$ (not depending on $u$, in the
first approximation), to be used for a fitting to the experimental
$P(a|u)$. This study is of interest and can be made elsewhere.

(vi) The marginal distribution $P_{\mathrm m}(a)$ is evaluated by
integrating out $u$,
\be\label{Pmarginal}
P_{\mathrm m}(a)=\int_{-\infty}^{\infty}\!\!\! {\mathrm d} u\,
P(a|\alpha(u),\lambda(u))g(u),
\ee
where in the first approximation the distribution $g(u)$ can be
taken Gaussian, and then should be compared with the experimental
unconditional distribution. This requires prior determination of
the dependence $\lambda(u)$ which can be made elsewhere.

We conclude by a few remarks.


(a) As it has been mentioned above, the distribution
(\ref{PLaval}) can be used "as is" for a fitting to the
experimental unconditional $P(a)$, as shown in Fig.~\ref{Fig1}. It
should be stressed however that in the proper RIN approach it is
the {\em marginal} distribution (\ref{Pmarginal}) obtained from
(\ref{PLaval}) by averaging over random noise intensities with
some judiciously chosen distributions assigned to them, that
should be fitted to the experimental unconditional $P(a)$. The
simplest choice is to assume that only $\alpha$ is a random
parameter, inverse of which follows chi-square or log-normal
distribution in the spirit of simple RIN models~\cite{Aringazin5}.
Particularly, for $\alpha=e^{u/u_0}$ the choice of the log-normal
distribution is equivalent to that $u$ is normally distributed
with zero mean~\cite{Aringazin4}. It should be emphasized here
that only absolute value of $u$ contributes to the marginal
distribution. In this case the distribution (\ref{PLaval}) is
treated as a conditional probability distribution function
$P(a|\alpha(u))$ (see Fig.~7 in Ref.~\cite{Aringazin5}), which
could be in principle fitted to the experimental conditional
distribution $P(a|u)$. Although we have found a good qualitative
agreement of the model with the experimental data on conditional
statistics, an illustrative quality of the representation of
experimental $P(a|u)$ in Ref.~\cite{Mordant0303003} does not allow
us to make a reliable {\em numerical} fit of the proposed
$P(a|u)=P(a|\alpha(u),\lambda(u))$. For good fit results, high
accuracy experimental data on $P(a|u)$ and on the contribution to
fourth order conditional moment, $a^4P(a|u)$, for $|u|$ ranging
from zero to three {\it rms} velocity with the step 0.5, would be
required.

(b) This would give a possibility to identify the dependence
$\lambda(u)$, for which one can try a polynomial or exponential
function.

(c) Despite similarity one observes some difference between the
experimental distributions $P(a)$ and
$P(a|u)$~\cite{Mordant0303003} which could be described in terms
of the RIN approach to LDN type model (\ref{PLaval}) along the
line of reasoning given in the present paper. Namely, with
$P(a|u)$ taken to be the distribution $P(a|\alpha(u),\lambda(u))$
and $P(a)$ to be derived from it by integrating out velocity
fluctuations, Eq.~(\ref{Pmarginal}).

(d) Reynolds-number dependence of the acceleration statistics is
not considered in the present paper and can be studied elsewhere.
Also, it is of interest to study variation of the value of
flatness factor of acceleration distribution measuring Lagrangian
intermittency as a function of finite time increment used for
low-pass filtering. This dependence exhibits a {\em fine
structure} of the viscous dissipation range of time-scales. The
flatness factor of the $R_\lambda=690$ flow varies by about 15\%
when the filter width is changed from $0.23\tau_\eta$ to
$0.31\tau_\eta$, where $\tau_\eta$ is Kolmogorov
time~\cite{Mordant0303003}.

\end{document}